\documentclass[twocolumn]{article}

\usepackage[l2tabu,orthodox]{nag}  
\usepackage[utf8x]{inputenc}       
\usepackage{upquote}               

\usepackage{graphicx}              %
\usepackage{amsmath}               
\usepackage{url}                   
\usepackage[caption=false]{subfig} 

\usepackage{verbatim, xcolor}
\usepackage{enumitem}
\usepackage{acronyms}
\usepackage[square,sort,comma,numbers]{natbib}

\usepackage{paralist}
\usepackage{whix} 
\usepackage{todonotes}

\newcommand{\superscript}[1]{\raisebox{1ex}{\ensuremath{{\mathrm{#1}}}}}
\def\we{\superscript{*}}
\def\wh{\superscript{\dag}}
\def\wg{\superscript{\ddag}}
\def\ws{\superscript{\natural}}

\usepackage{hyperref}
\hypersetup{colorlinks=true,allcolors=royalblue}
\usepackage{breakurl}
\begin{document}






%

\title{RemIX: A Distributed Internet Exchange\\for Remote and Rural Networks}
\author{
    William Waites\we\wh \and James Sweet\we\wh \and Roger Baig\wg \and
    Peter Buneman\we \and Marwan Fayed\ws \and Gordon Hughes\we \and
    Michael Fourman\we \and Richard Simmons\ws
}

\maketitle

\let\thefootnote\relax\footnote{
  \noindent {\we}University of Edinburgh
}
\let\thefootnote\relax\footnote{
  \noindent {\wh}HUBS \textsc{c.i.c}
}
\let\thefootnote\relax\footnote{
  \noindent {\wg}guifi.net
}
\let\thefootnote\relax\footnote{
  \noindent {\ws}University of Stirling
}
\let\thefootnote\relax\footnote{
  \noindent Corresponding authors: \url{wwaites@tardis.ed.ac.uk},
  \url{opb@inf.ed.ac.uk}, \url{ mmf@cs.stir.ac.uk}
}

\begin{abstract}
The concept of the \ac{IXP}, an Ethernet fabric central to the structure of the
global Internet, is largely absent from the development of community-driven
collaborative network infrastructure. The reasons for this are two-fold.
\acp{IXP} exist in central, typically urban, environments where strong network
infrastructure ensures high levels of connectivity. Between rural and remote
regions, where networks are separated by distance and terrain, no such
infrastructure exists. In this paper we present RemIX a distributed \acp{IXP}
architecture designed for the community network environment. We examine this
praxis using an implementation in Scotland, with suggestions for future
development and research.

\end{abstract}

%
%

\section{Introduction} \label{sec:intro} In remote and rural regions the last-mile problem has been the subject of much
focus. Deployments in remote regions of the world have shown that it is possible
to build high quality access networks in otherwise under-serviced
regions~\cite{guifi, tegola,Hasan:2015}. Their underlying technologies range in medium (eg. copper
or fibre-optic cabling, licensed or unlicensed wireless), energy (eg. solar or
wind generation or mains supplied), and topology.
Successful deployments, including our own in Scotland,
have two attributes in common:
\begin{inparaenum}[(i)]
  \item Networks designs are bespoke, suggesting
    there is no one-size-fits-all solution;
  \item crucially, communities must be invested and
    involved~\cite{Wallace:2015a,Wallace:2015b}.
\end{inparaenum}

Though remote access network research is far from complete, the next question is
increasingly clear: What options do remote, isolated networks have for
`backhaul' to interconnect with the rest of the Internet? We define ``remote''
as far from urban areas where commodified network infrastructure is available.
For example long-distance circuits, if and where they are available, are both
expensive and difficult to reach. Access networks in remote places serve
populations that are dispersed. The lower population density reduces the size of
their user-base when compared to their urban cousins. With no options for
interconnecting with nearby networks to generate economies of scale,
\emph{high-quality} backhaul is prohibitively expensive, if it exists at all.


The absence of resource pooling options for remote networks is the focus of this
paper. One such example is operated by the Guifi Foundation~\cite{guifi}. Guifi
operates a regional backbone network as a commons. The abstraction that is
presented to clients is an exchange point implemented over IP. In this type of
network, relationships between end-users are either mediated by Guifi, or
implemented as an overlay.

The \acf{IXP} is a long-standing structure that plays a pivotal role in
facilitating interconnections between networks~\cite{Ager:2012,Chatzis:2013}. We
are motivated by \acp{IXP} for two reasons. First, the primary role of an
\ac{IXP} is economic. Member networks can connect $n$ networks at an IXP with
$n$ circuits, rather than arranging $O(n^2)$ circuits independently.
Second, the \ac{IXP} model of multilateral public peering leads to high density
interconnections, and traffic across the exchange that can be comparable in
magnitude to the largest global service providers~\cite{Ager:2012}. Together,
they are an indication that such a topology might be used to improve
inter-connectivity between networks in under-serviced regions, and to pool
otherwise expensive backhaul resources.

In this paper we present RemIX, a \emph{distributed} Internet Exchange for
Remote and rural networks. The RemIX architecture is agnostic to underlying
technologies, embedding the same principles as the successful remote networks it
is designed to serve. It distinguishes itself from \acp{IXP} by the vast
distances permitted between points of presence, and the lower density of member
networks that connect to them. The trade-off between distance and density gives
rise to the idea of \emph{lightweight} points of presence. The lightweight
nature is advantageous, in that as few as two member networks are sufficient to
establish a point of presence.

We describe our RemIX implementation in Scotland. In its current form our
deployment services a $\sim 2000\mathrm{km}^2$ region that spans sea and mountainous
mainland. Implementation details are provided, with motivating rationale, so
that others may benefit from our efforts. Functionally, our implementation
appears to its members as a large Ethernet switching fabric. Crucially, RemIX
allows member networks to establish unmediated relationships between themselves.

In the following sections we further motivate \acp{IXP} as an ideal model. We
then discuss the RemIX architecture in detail. Our deployment is described,
along with lessons learned. Finally, a broader context of the local environment
is presented before concluding remarks.

\section{\aclp*{IXP}} \label{sec:context} 


As part of the decommissioning of the \acs{NSFNET},
four \acp{NAP} were created. They were operated by large
American telephone companies (MCI, Sprint, PacBell, Ameritech) and
designed to prevent partitioning of the commercial
Internet~\cite{Ager:2012,Chatzis:2013}.
The \acp{NAP} were prohibitively expensive and had arbitrary
technological requirements which created barriers to participation.
Soon \acp{IXP} emerged as an alternative. \acp{IXP} appeared in
carrier-neutral facilities allowing dense inter-network connections on
a non-discriminatory basis. Presence at an \acp{IXP} entails freedom
to make bilateral arrangements with any other network also
present. Worldwide, \acp{IXP} now number in the hundreds and are a
fundamental feature in the structure of the Internet.



A mirroring of this structure would be useful in joining remote
networks. The increases of interconnection density could then
be used to pool traffic, and make collective use of expensive resources
such as long-distance circuits. However, there are some important
differences between the environment of a typical urban \ac{IXP} and
the rural regions, as in the West Coast of Scotland:
\begin{inparaenum}[(i)]
  \item There are no data centres, carrier-neutral or otherwise;
  \item due to geography there is no single facility where all of the
    networks could meet.
\end{inparaenum}

\section{RemIX Architecture} \label{sec:arch} In this section we present the RemIX architecture. We compare RemIX with IXP architectures, and relate those
benefits in the context of remote access networks.

\subsection{Design Requirements}

Our requirements are shaped by three broad goals:
\begin{inparaenum}[(i)]
  \item establish high-quality backhaul to remote regions;
  \item ensure backhaul affordability for small access networks;
  \item allow networks to maintain the autonomy that is
    fundamental to their sustainability.
\end{inparaenum}
Member networks must be able to connect to one or more transit providers.
Members must also be free to arrange and articulate policies among themselves.
These requirements imply that a \emph{logical} concentration of inter-network
connections is desirable, which suggests a shared switching fabric below the
network layer.

\begin{figure*}
  \subfloat[Traditional IXP]{
    \resizebox{0.45\columnwidth}{!}{
      \begin{tikzpicture}
        \ixboxesA
      \end{tikzpicture}
      \label{fig:ixbA}
    }
  } \hfill
  \subfloat[Modern urban IXP]{
    \resizebox{0.45\columnwidth}{!}{
      \begin{tikzpicture}
        \ixboxesB
      \end{tikzpicture}
      \label{fig:ixbB}
    }
  } \hfill
  \subfloat[RemIX]{
    \resizebox{0.45\columnwidth}{!}{
      \begin{tikzpicture}
        \ixboxesC
      \end{tikzpicture}
      \label{fig:ixbC}
    }
  }
  \caption{Comparison of exchange point models. Notice density.}
  \label{fig:ixb}
\end{figure*}
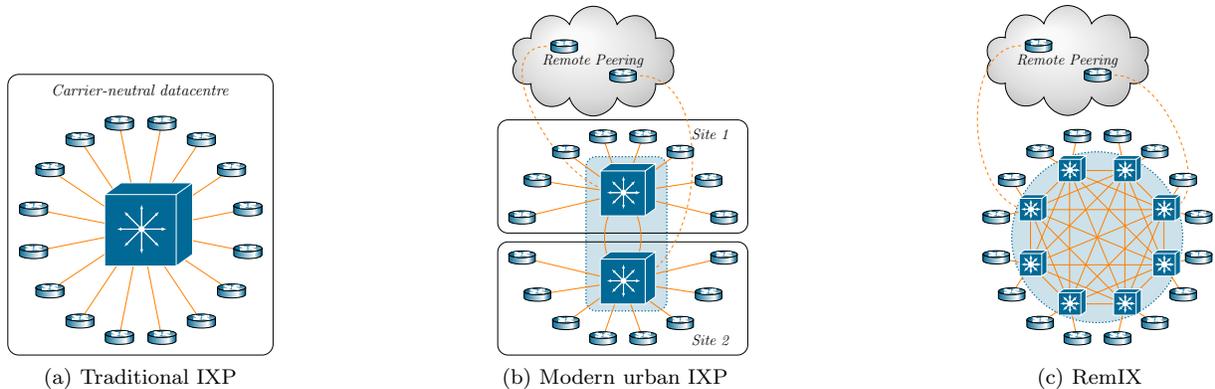

Networks that are capable of connecting to the same location can do so with an
Ethernet switch. This is the basis for traditional \ac{IXP} design (Fig.~\ref{fig:ixbA}) where member networks connect to a central fabric
with their own router that sits inside the IXP facility. Our remote networks
have no such luxury. In response, we take and distribute the contemporary design
of a multi-site \ac{IXP} (Fig.~\ref{fig:ixbB}). A multi-site \acp{IXP}
presents a single logical fabric to its members, implemented with switches
that are joined by private circuits.

The RemIX architecture that emerges (Fig.~\ref{fig:ixbC})
has no large facility nor physical housing. Instead it is
distributed so that \emph{lightweight} points of presence may be established
where there are as few as two members. Members either colocate their border
routers with the exchange switch, or remotely on the far end of a link, as
circumstances dictate.

These circumstances motivate the lightweight nature of points of presence.
Since the fabric is distributed, fewer networks that will connect from each
site. High port densities are unnecessary. Simultaneously, space and power are
both at a premium. For example, a remote port into RemIX could be housed in a
small cabinet atop a hill, or in space that is donated by a property owner for
this purpose. Equipment is therefore restricted to the small and
power-efficient.

\subsection{RemIX Components}

\subsubsection{Switching Fabric}

The exchange itself must mimic a distributed Ethernet switch. Multiple
Ethernet-like link options include fibre, 802.11 wireless, licensed wireless, fibre, leased pseudo-wires. The switching fabric may be implemented on
top using \acs{BGP}-\acs{VPLS}~\cite{rfc4761} (as we have in
Section~\ref{sec:bgpvpls}),
\acs{BATMAN}~\cite{johnson2008simple}, or
\acs{TRILL}~\cite{perlman2004rbridges} protocols. The salient feature
between them is MAC address learning to establish an Ethernet switch
similar to the \ac{MEF} E-LAN interface
specification~\cite{mef62}.

\subsubsection{Member \acp{AS}}

Among traditional \acp{IXP} connected networks are encapsulated into
Autonomous Systems (\acp{AS}). Among RemIX member networks, the
policies of the small sized member networks are
different from the Internet's \ac{DFZ}. In particular, member
networks' smaller routers will be neither be capable of storing the
entire Internet routing table, nor are they likely to announce
netblocks large enough to be globally visible.  However, \ac{AS}
\emph{encapsulation} enables networks to retain their internal
structures and methodologies, and to interconnect safely  with
neighbours. Due to the likelihood of collisions use of private
\acp{ASN} is inappropriate for this purpose~\cite{rfc6996}, as are
private IP addresses for the exchange itself~\cite{rfc1918}.

\subsubsection{Exchange Transit}

RemIX members' IP address spaces will be small, and need some entity to
advertise larger netblocks on their behalf. This suggests a specialized transit
provider to mediate between members and the wider Internet. For this reason
RemIX members form a confederation with a transit provider that
presents them collectively to upstream providers and other exchange points. This
is unusual for \acp{IXP}: Rarely are transit relationships implemented with
exchange points. However, this is normal in RemIX, and likely necessary to
function in the intended environment. We note that transit service should be
optional to members, with no requirement to purchase said provider's transit as
a condition for joining the exchange. Also, nothing prevents other such
providers from participating.

\subsubsection{Auxilliary Services}

\acs{BGP} configuratoin can be a complex. For example, upon
connecting to RemIX, member networks need to be configured to peer amongst
themselves. The complexity quickly increases as session numbers grow with the
square of the number of participants. Instead, \acp{IXP} use
\emph{route-servers} to repeats announcements from one member to all others. A
route reflector keeps the configuration burden to a minimum. Other useful
services such as \acs{NTP} clocks and looking glasses for assistance in
debugging may be offered in addition.


The overall RemIX architecture is motivated by our own needs in Scotland. In
the next section we present our first-phase implementation of RemIX,
alongside remarks on usability and directions for the future.


\section{RemIX Deployment in Scotland} \label{sec:impl} 

In this section we describe our first implementation of RemIX in a series of planned deployments across Scotland. In the West Highlands there
is a cluster of
11 small community networks. Their spread across $\sim 2000km^2$
of sea and mountainous islands makes the
construction of an exchange fabric geographically ambitious. Four networks have
a history of interconnecting and sharing network resources, pre-established
relationships that must be respected in our deployment.

Our deployment's location is its namesake, the \ac{WHIX}. Both logical
and physical layers are described below, with additional lessons and
comments drawn from our experience.

\subsection{West Highland IX at Layer 1}

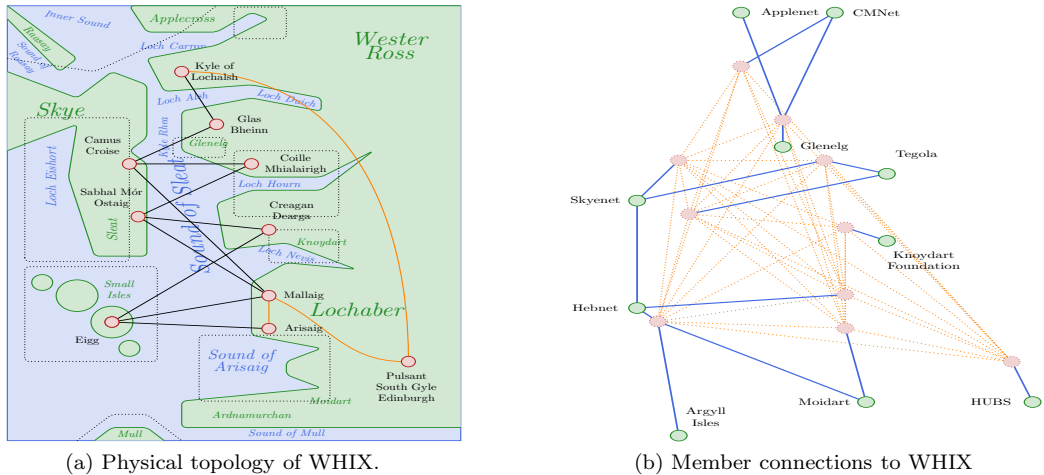
\begin{figure*}
  \centering
  \subfloat[Physical topology of \ac{WHIX}.]{
    \resizebox{0.8\columnwidth}{0.3\textheight}{
      \begin{tikzpicture}
        \whixphysicaldiagram
      \end{tikzpicture}
    }
    \label{fig:whixmap}
  }\hfil 
  \subfloat[Member connections to \ac{WHIX}]{
    \resizebox{0.8\columnwidth}{0.3\textheight}{
      \begin{tikzpicture}
        \whixmeshdiagram
      \end{tikzpicture}
     }
    \label{fig:phytop}
  }
  \caption{Physical and logical layout of \ac{WHIX}. In
  Figure~\ref{fig:whixmap} the dark lines correspond to radio links
  and the light, curved lines to leased ethernet circuits.
  In Figure~\ref{fig:phytop} the dashed lines
  correspond to internal layer-2 circuits forming \ac{WHIX}
  switching fabric and the solid lines to member connections.}
\end{figure*}

The physical \ac{WHIX} fabric is overlayed onto a stylized map of the region in
Figure~\ref{fig:whixmap}. The map itself preserves critical geographical
features. Red connected nodes are the connection sites. In a traditional
\ac{IXP} these sites are the Ethernet ports into which subscriber \acp{AS}
plug-in. WHIX sites are connected by wireless radio links in black, and leased
100Mbps or 1Gbps circuits in orange. The areas enclosed with dotted lines
correspond to the service areas reachable from each site.

We complement \ac{WHIX}' physical topology in Figure~\ref{fig:whixmap} with
the member network in Figure~\ref{fig:phytop}. In the latter, unlabeled red
nodes are the \ac{WHIX} points of presence and correspond with the same set of
red nodes in Figure~\ref{fig:whixmap}. The dashed lines represent the fully
connected virtual topology that implements the exchange E-LAN.

The two places in the region where long-distance ethernet circuits are
available on the mainland are the towns of Mallaig and Kyle of
Lochalsh. Circuits\footnote{At the time of writing, the circuit from
Mallaig is in place, and that from Kyle is planned.} from these sites
connect back to the Pulsant datacentre in Edinburgh to facilitate
remote peering --- and indeed the provision of Internet access via the
exchange point.

The radio links are implemented with equipment from Ubiquiti Networks,
configured in transparent bridge mode so that they appear as Ethernet from a
functional perspective. The switching fabric itself at each of \ac{WHIX} points
of presence is implemented with Mikrotik routers. This choice was made because
of their moderate port density, low power consumption, low cost, and adequately
featureful \acs{MPLS} implementation. We revisit this choice in the next
section. All equipment is configured to pass Ethernet frames of at least 1600
bytes to provide room for the necessary extra protocol headers for implementing
the E-LAN service.

\subsection{West Highland IX at Layer 2}\label{sec:bgpvpls}

We emphasize that layer-2 details are \emph{internal} to
\ac{WHIX}, and invisible to members who only see an Ethernet switch. Also, our
implementation decisions are by no means the only possible means of
implementation.

In \ac{WHIX} the requirement for functional equivalence to a \acs{MAC} address
learning Ethernet switch is met using \acs{BGP} signalled
\acs{VPLS}~\cite{rfc4761}. This creates a full set of \acs{LSP} pseudo-wires
between every pair of \ac{WHIX} edge routers.
Each \ac{WHIX} router maintains an \acs{OSPF} routing protocol
adjacency with its neighbours and distributes reachability information
for its loopback IP address. All addresses used for this purpose are
private IPv4 addresses~\cite{rfc1918}. This is the basic layer that
ensures reachability throughout the distributed fabric.
Non-IP traffic is carried via \acs{LDP}~\cite{rfc5036}
with \acs{MPLS} labels according to the topology of the underlying \acs{OSPF}
network.

Routers in \ac{WHIX} establish \acs{BGP} peering sessions with routers at
Mallaig and Sabhal M\`{o}r Ostaig that act as route reflectors~\cite{rfc4456}.
Participating routers use route reflectors to exchange reachability information
without requiring a full mesh ($n^2$) of internal peering sessions. The presence
of \acs{BGP} signalling throughout the \ac{WHIX} fabric enables the use of
multi-protocol extensions~\cite{rfc4760}. Routers can use extensions to signal a
desire to form part of the exchange \acs{LAN}.
The result is a fully meshed \acs{VPLS}, where each router has a virtual bridge
interface that forms part of the exchange \acs{LAN}.

Interfaces can be added to virtual bridges, as needed, to
form part of the exchange. Care must be taken to prevent loops in which members
see the traffic that they originate. This is accomplished with a split-horizon
method~\cite{rfc4762}. Equally, members must be prevented from creating bridge
loops via their own network by employing \acs{MAC} address filter on relevant
ports.

\subsection{West Highland IX at Layer 3+}

Given logical connectivity between all member ports, it remains to assign IP
addresses to their border routers, as well as public infrastructure such as the
router server. As mentioned above the use of private IP address space for this
purpose is undesireable since it generates risks of collisions with members' own
infrastructure. \ac{WHIX}, and more generally RemIX, is fortunate in this
regard: The design meets the definition of an \ac{IXP}~\cite{ripe451,whixrules},
making it possible to acquire IPv4 and IPv6 address allocations from RIPE
NCC~\cite{ripe649}.

\begin{figure}[h]
  \resizebox{0.8\linewidth}{!}{
    \begin{tikzpicture}
      \whixtopodiagram
    \end{tikzpicture}
  }
  \caption{
  Autonomous System topology. The members of a RemIX form a fully
  connected network where each may communicate with another over the
  exchange without intermediation.
  }
  \label{fig:l3}
\end{figure}
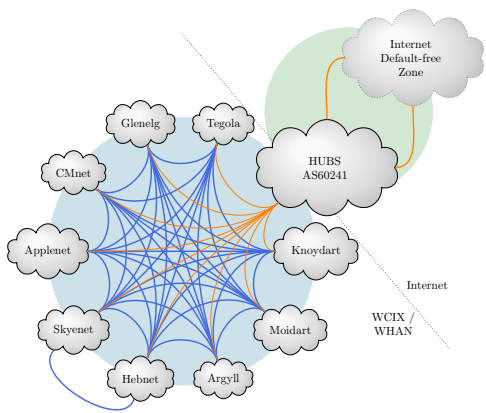

The full layer-3 \ac{WHIX} topology is shown in Figure~\ref{fig:l3}. At this
stage member network have everything they need. Members can communicate at layer
2. Each has an IP address at layer 3, an autonomous system number for
identification, and their own networks to announce. Bilateral peering
arrangements (an otherwise $n^2$ configuration task) are facilitated by two
route-servers, as before at Mallaig and Sabhal M\`{o}r Ostaig. The route servers
redistribute reachability information, akin to a route-reflector omits its own
\ac{ASN} from the path.

The transit provider, HUBS (see Section~\ref{subsec:hubs}), is also
present at \ac{WHIX} as a member. In addition to public multilateral peering, it
establishes bilateral sessions with members wishing to announce either a
default route or full Internet routing tables. HUBS forwards those members'
announcements upstream and to their peers. In this way transit, and hence
connectivity to the global Internet, is provided over the exchange.

\subsection{Deployment Discussion}

Our experience motivates higher-level comments to further distinguish
RemIX deployments form their larger \ac{IXP} cousins. Flat networks
consisting of a single layer-2 broadcast domain can be plagued by
problems that are difficult to troubleshoot. By its very design RemIX requires
that members be able to communicate directly without mediation at the
IP layer. Like other \acp{IXP} RemIX eliminates a large class of
potential problems by allowing only unicast and \acs{ARP} traffic on
the exchange. Moreover, members must nominate a specific \acs{MAC}
address for their connections, which reduces the risk of loops and
broadcast storms. We also adopt best practices such as quarantines for
new connections while they are evaluated for correctness.



IP transit in RemIX also deserves to be addressed. Transit
via the exchange, for networks that are not otherwise visible on the Internet,
may evoke notions of conflicting interests that
beset \acp{NAP}. However the similarity is superficial. Here, the transit
provider and exchange operator HUBS \textsc{c.i.c.}, is a cooperative
that exists for the benefit of and is controlled by the members, who
are also members of the \ac{IXP}. As a consequence all parties'
economic interests are aligned.

Finally, \ac{WHIX}' implementation using \acs{BGP}-\acs{VPLS}
to construct the exchange fabric makes it possible to offer auxiliary
point to point pseudo-wire services to its members. This is
useful for those members that have need for making connections internal to their
networks.


\section{The Environment}\label{subsec:hubs}

Scotland contains 1/3 of the area but 10\% of the population of Great Britain. It also has 95  inhabited islands with about
 100,000 people. The Scottish Highlands and Islands, where this
work is currently focussed, consist of mountainous terrain stretching along a 400km north to south corridor. Islands are scattered on the West together with
deep lakes and glens penetrating the mainland to the East.  The economy was
traditionally maritime, and nearly all habitation is at sea level or in the
glens.

Fibre in the region has only recently appeared.  Much of the telephone network
in the region was constructed with microwave links. Infrastructure is improving,
though plans terminate at telephone exchanges. Among them, fibre-based
services are rare. In the medium term future, local wireless distribution is the
only feasible technology for adequate bandwidth and quality of service.

\begin{figure}[h]
\centering
 \includegraphics[width=0.8\columnwidth]{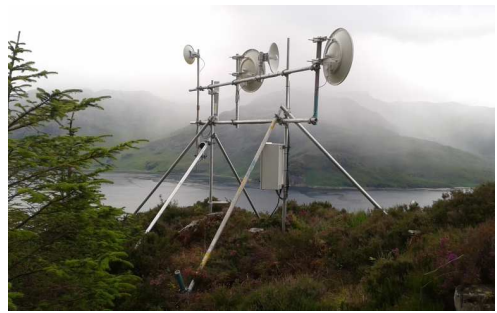}
 \caption{A basic relay}
\label{fig:mhialairigh}
\end{figure}

Starting in 2008, the Tegola project~\cite{tegola} started to
experiment with technology that would enable communities to build
their own wireless networks.
The details of Tegola, and its dissemination to nearby communities, are
omitted due to space constraints. Relevant to this project is the technology
that emerged. Figure~\ref{fig:mhialairigh}, for example, features the type of
robust, inexpensive relay construction that operates in mountainous region, and
that can be constructed by its residents.

Many communities have since constructed their own local distribution networks
with point-to-point wireless links that can span more than 20km. Expertise is
often shared between them, also infrastructure where feasible, yet they operate
independently. Constrained by availability, they acquire backhaul via ADSL lines
nearby to telephone exchanges. Ethernet services have since emerged in two
larger towns, with wholesale pricing that exceeds the budget of any single
community. A resolution has two components: An organizational vehicle that
combines networks to generate economies of scale, and a supporting network
infrastructure.


We have learned that solutions are complicated by both terrain and by culture.
In particular we note:
 (a) Social aspects and organization of communities can fail to align with the
  ideal ``electronic'' or networked communities, eg. physical landscape constrains
  connectivity, while social and economic groups can be determined by vehicles for funding.
(b) Local network infrastructure is non-uniform and varies in complexity.
(c) Communities that share network resources generally do so in a
  non-systematic or ad-hoc manner.


\subsection{HUBS C.I.C.}\label{subsec:hubs}

In response to the local environment and absence of affordable
backhaul, the Universities of Edinburgh and Stirling launched
HUBS, which is a not-for-profit transit provider whose members are the
community networks that it serves.  HUBS is also a co-operative where
the networks that subscribe also become members. It is the culmination
of collaborations between Universities with communities in the West
Highlands, and later with community networks in the South Scotland.

The need for RemIX-like functionality arose soon after launch. Two of
the subscriber networks took advantage of mutual proximity to
collaborate on an operational basis. Equipment management and
troubleshooting tasks, for example, were shared. Their desire to keep
the details internal was complicated by the fact that their only
interconnection was mediated by HUBS. Circuits were hand-crafted
between them, and demonstrated the benefits of bilateral agreements
between HUBS members. However, while effective, hand-crafted circuits
would fail to scale.

HUBS bridges gaps in backhaul affordability. It has also revealed the
benefits emerge when remote and rural networks are able to act
collectively in the wholesale telecommunications market, and present a
uniform interface to their upstream transit provider. However, a
transit-only solution prohibits autonomous arrangements between
members unmediated at the IP layer. From this need the RemIX
architecture emerges.

\section{Concluding Remarks} \label{sec:conc} 

The features of RemIX described above will be instantly recognizable to anyone
who has participated in a regular \ac{IXP}. This is by design. RemIX is
architected to mirror in under-serviced regions, the benefits of \acp{IXP} in
urban regions. The encapsulation of small community networks in \acp{AS} means
that they can present a uniform interface to a transit provider, cooperate and
share resources. RemIX provides these benefits to members without sacrificing
their independence, a necessary attribute for longterm sustainability.
%

\section{Acknowledgements}
This project is supported by the University of Edinburgh, by the
Scottish Government and by local industries including Marine Harvest
Scotland and Benchmark Holdings.  During an incubation period,
\ac{WHIX} is jointly operated by HUBS and the University of Edinburgh.

\bibliographystyle{abbrv}
\bibliography{paper_cix}

\end{document}